\documentclass[acmsmall]{acmart}
\AtBeginDocument{%
  \providecommand\BibTeX{{%
    \normalfont B\kern-0.5em{\scshape i\kern-0.25em b}\kern-0.8em\TeX}}}

\acmConference[CSCW '24]{November 09--13, 2024}{San Jose, Costa Rica}

\setcopyright{acmlicensed}
\acmJournal{PACMHCI}
\acmYear{2024} \acmVolume{8} \acmNumber{CSCW2} \acmArticle{496}
\acmMonth{11}\acmDOI{10.1145/3687035}

\usepackage{rotating}
\usepackage{CJKutf8}
\usepackage{xcolor}
\usepackage{colortbl} 
\usepackage{booktabs}  
\usepackage{arydshln}  
\usepackage{tabularx}  
\usepackage{lipsum}  
\usepackage{hhline}
\usepackage{makecell}
\usepackage{array}
\usepackage[normalem]{ulem}

\definecolor{DarkGreen}{HTML}{5DAC81}
\definecolor{DarkOrange}{RGB}{237, 125, 49}
\definecolor{myblue}{RGB}{173, 216, 230}

\newcommand\review[1]{\textcolor{black}{#1}}
\newcommand\CSCWreview[1]{\textcolor{black}{#1}}

\begin{document}



\title[Parental Needs for AI-based Storytelling]{Exploring Parent's Needs for Children-Centered AI to Support Preschoolers' \CSCWreview{Interactive} Storytelling and Reading Activities}


\author{Yuling Sun}
\affiliation{%
  \institution{East China Normal University}
  \country{China}}
\email{ylsun@cs.ecnu.edu.cn}

\author{Jiaju Chen}
\affiliation{%
  \institution{East China Normal University}
  \country{China}}
\email{10205102450@stu.ecnu.edu.cn}

\author{Bingsheng Yao}
\affiliation{%
  \institution{Rensselear Polytechnic Institute}
  \country{USA}}
\email{yaob@rpi.edu}

\author{Jiali Liu}
\affiliation{%
  \institution{East China Normal University}
  \country{China}}
\email{51212500056@stu.ecnu.edu.cn}

\author{Dakuo Wang}
\authornote{Corresponding author}
\affiliation{%
  \institution{Northeastern University}
  \country{USA}}
\email{d.wang@northeastern.edu}

\author{Xiaojuan Ma}
\affiliation{%
  \institution{Hong Kong University of Science and Technology}
  \country{Hong Kong SAR}}
\email{mxj@cse.ust.hk}

\author{Yuxuan Lu}
\affiliation{%
  \institution{Northeastern University}
  \country{USA}}
\email{lu.yuxuan@northeastern.edu}

\author{Ying Xu}
\affiliation{%
  \institution{Harvard University}
  \country{USA}}
\email{ying_xu@gse.harvard.edu}

\author{Liang He}
\affiliation{%
  \institution{East China Normal University}
  \country{China}}
\email{lhe@cs.ecnu.edu.cn}
\renewcommand{\shortauthors}{Yuling Sun, et al.}

\begin{abstract}

\review{
Interactive storytelling is vital for preschooler development. While children's interactive partners have traditionally been their parents and teachers, recent advances in artificial intelligence (AI) have sparked a surge of AI-based storytelling and reading technologies. 
As these technologies become increasingly ubiquitous in preschoolers' lives, questions arise regarding how they function in practical storytelling and reading scenarios and, how parents, the most critical stakeholders, experience and perceive these technologies. 
This paper investigates these questions through a qualitative study with 17 parents of children aged 3-6. 
Our findings suggest that even though AI-based storytelling and reading technologies provide more immersive and engaging interaction, they still cannot meet parents’ expectations due to a series of interactive and algorithmic challenges. We elaborate on these challenges and discuss the possible implications of future AI-based interactive storytelling technologies for preschoolers.  
}
\end{abstract}



\received{January 2024}
\received[revised]{April 2024}
\received[accepted]{May 2024}

\maketitle

\section{Introduction}

\CSCWreview{
\textit{Interactive storytelling} is a specific form of storytelling where, in addition to merely narrating the story verbatim, parents often sit together with and read to their preschool children\footnote{``Preschoolers'' or ``preschool children'' in this paper refers to young children, typically aged 3 to 6, who are in the stage of early childhood education that precedes formal schooling~\cite{Preschooler} about the story content.}, interact with preschoolers and engage them in guided conversation~\cite{zevenbergen2003dialogic} by asking questions and providing responsive feedback. 
It has been proven to be beneficial to the later language and literacy proficiency of preschoolers, communication skills, cultural and emotional awareness, and other aspects of cognitive development~\cite{wright, yen2018joint, chang2016mother, peck1989using}. }
However, due to various practical challenges, today's parents may not always have sufficient language skills, knowledge, time, or inclination to engage in such conversation-rich storybook reading with their \review{preschoolers}, even though they believe in the importance of interactive storytelling with their children
\cite{manz2010descriptive, xu2021same}.

In recent years, AI-based tools and systems with multiple functions, modalities, and interaction techniques have become increasingly popular in supporting \CSCWreview{interactive storytelling practices} for \review{preschoolers}. These tools can be applications on mobile phones or tablets \cite{storycoder, Storybuddy,zhao2022educational}, or exist with physical anthropomorphic tangible presences (e.g., Alpha Egg \cite{egg}, Luka \cite{luka}, Codi \cite{codi}), relying on interactive technologies tailored to children (e.g., \cite{FairytaleQA, yao2022ai}), or general voice assistant technologies (e.g., Amazon Alexa, Google Assistant, Apple Siri). 
\review{Some} of these tools are mature commercial products (e.g., Alpha Egg \cite{egg}, Luka \cite{luka}, Codi \cite{codi}) and some others are still in the research period (e.g., StoryCoder \cite{storycoder}, and StoryBuddy \cite{storycoder, Storybuddy}). 
With increasingly powerful and intelligent capabilities, these AI-based systems and tools have provided tremendous promises for interactive, intelligent, and engaging \CSCWreview{storytelling and reading experiences} for children, acting as their language partners or learning assistants \cite{xu2021same, storycoder}. Recent boom in large language model (LLM)-based technologies \cite{llm} further solidify this vision~\cite{mahmood2023llm}. This increasing functionality, diversity, and popularity of AI-based technologies in \CSCWreview{interactive storytelling }scenarios for \review{preschoolers} raise critical questions regarding: \textit{how AI techniques support real-world \CSCWreview{interactive storytelling and reading practices }and how parents, the vital role of storytelling practices, experience, and perceive AI-mediated storytelling for preschoolers?} 

In CSCW, studies have paid attention to the design, development, and use of various storytelling applications, with the aim of improving children's digital literacy \cite{storycoder, PlushPal}, stimulating their creativity and collaboration \cite{storycoder, TellTable}, fostering their connection and interaction with parents \cite{SignBright, Storybuddy}, etc. Meanwhile, researchers have evaluated the functionality and effectiveness of existing commercial AI products (e.g., Alpha Egg \cite{egg}, Luka \cite{luka}, Codi \cite{codi}) as well as the recent LLM technologies \cite{AIGC2023design,chen2023fairytalecqa} in supporting storytelling practices with \review{preschoolers} \cite{xu2021same, garg2022last, chubb2022interactive}. 
However, these studies were conducted primarily through experimental research. To date, very little has been published on the usage and evaluation of AI-based storytelling and reading technologies \CSCWreview{in real-world interactive storytelling scenarios} beyond a limited research period. 

More importantly, while these studies have paid attention to accessing the effectiveness and usability of AI technologies in children's growth and development, the experience, perceptions, and expectations of parents, who actually play the most significant roles in practical storytelling \cite{parentrole1, parentrole2} with the children, have been largely overlooked. In particular, given the cognitive and social development stage, \review{preschoolers} often lack the essential AI literacy \cite{su2023artificial} and skills \cite{lever2011discussing, resnick2009scratch} to independently use various AI technologies, thus relying on parents to navigate this process \cite{Storybuddy}. Neglecting the understanding of parents' needs and perceptions may significantly hinder the effectiveness of these technologies in real-world contexts.

Our study fills this research gap through a qualitative study of 17 parents with preschoolers between the ages of 3 and 6. Our findings indicate that \CSCWreview{parents in our study }have a cautiously positive attitude towards \CSCWreview{AI-based interactive storytelling technologies}. 
Given the practical challenges facing today’s parents in carrying out high-quality and meaningful interactive storytelling, our findings suggest that parents expect AI technologies to help alleviate their burdens and support the desired parent-children interactive storytelling. 
Yet, existing tools generally fail to support parents' expected storytelling in the real world due to several interactive and algorithmic challenges. 
Drawing on our findings, we delve into parents' expectations for ideal parent-child interactive storytelling and propose several crucial design implications for future research. Our study contributes to the CSCW community by 1) acquiring an empirical understanding of parents' perception and facing challenges regarding the use of AI-based storytelling tools in practice through a qualitative study involving 17 parents with preschoolers, and 2) identifying and analyzing vital design implications and recommendations to better facilitate the increasing demand for AI-based children-centered storytelling.



\section{Related Work}
Our study situates itself in the research of AI-based storytelling for \review{preschoolers} and its practical effectiveness. In this section, we first present a review of existing AI-based storytelling tools. Following that, we review existing literature about the practical usage and evaluation of these tools from children's and parents' perspectives, respectively. 

\subsection{AI-Based Interactive Storytelling Systems}

With increasingly powerful and intelligent capabilities, as well as the practical barriers and challenges in traditional parent-child storytelling, there has been a growing interest in CSCW and HCI in designing and using AI-based technologies in children's storytelling scenarios \cite{frohlich2009storybank, storycoder, PlushPal, TellTable, SignBright, Storybuddy, rubegni2014fiabot}. Researchers, for instance, have designed StoryCoder \cite{storycoder}, a voice-guided smartphone application, and PlushPal \cite{PlushPal}, a web-based tool, to improve children's computational thinking ability and data literacy by leveraging storytelling as a creative activity. \citet{TellTable} designed TellTable, a tabletop storytelling application, to stimulate children's creativity and collaboration.
\citet{SignBright} and \citet{Storybuddy} designed SignBright and StoryBuddy, respectively, with the aim of fostering connection, collaboration, and understanding between children and parents in storytelling scenarios. 
\citet{EmotionBlock} designed EmotionBlock, a tangible toolkit for children-oriented social-emotional learning through storytelling. 
Meanwhile, researchers have paid specific attention to children with disabilities, such as \citet{Wildcard}'s Wildcard for children with intellectual developmental disabilities and \citet{SignBright}'s SignBright for deaf children. Besides, commercially available AI-based robotic toys, such as Alpha Egg \cite{egg}, Luka \cite{luka}, Codi \cite{codi}, etc., also provide platforms for facilitating interactive storytelling experiences for children \cite{garg2020conversational}.

The recent boom of Large Language Model (LLM) technologies further sparked the potential of AI-based technologies, particularly with a verbal communication interface \cite{mahmood2023llm, yang2024wish, wan2024building, wu2024sunnie}, in supporting more intelligent and personalized storytelling practices. \citet{AIGC2023design}, for instance, examines the design implications of leveraging generative AI tools to improve children's literacy development and creative story expression. Based on the findings, they proposed AIStory \cite{han2023aistory}, a prototype of a generative AI-powered visual storytelling application that can be used for children’s creative expression, storytelling, and literacy development. \citet{li2023designing} propose a peer-like embodied conversational agent named STARie that integrates multiple AI models
that aims to support narrative production in collaborative storytelling, specifically for children aged 4-8.

While these studies provide the worth-expecting vision for AI-mediated storytelling, most existing technologies are still in the prototype design and evaluation phase, involving brief testing for a short time with limited users rather than applying them to real-world scenarios for a long time. Due to the lack of large-scale deployment and long-term usage, it remains unclear how such AI-based storytelling systems are perceived and used by their intended users (e.g., parents and \review{preschoolers}) and what barriers and challenges they might encounter (e.g., ethical concerns~\cite{isaza2023fairy} or trust in AI issues~\cite{drozdal2020trust}). 
In light of the trend of AI-mediated storytelling for \review{preschoolers}, urgent attention is needed to deeply understand these aspects, which can help the designers and developers of AI-based storytelling systems better design these technologies integrate into the practical storytelling scenario to form a ``human-parent collaboration'' team~\cite{wang2019human,wang2020human}. 
Our study contributes to bridging this gap by examining the use of existing AI-based storytelling tools in the real world, particularly from parents' perspective.

\subsection{AI-Based Storytelling Technologies and Children}

In CSCW and related fields, there has been a surging interest in evaluating the performance of AI-based intelligent systems in children’s reading and storytelling scenarios. Researchers, for instance, have illustrated that preschool-age children can generally interact naturally with conversational agents (CAs), perceiving them as intelligent and friendly \cite{druga2017hey}.
Interacting with CAs can also improve children’s story comprehension and learning \cite{xu2021same, kory2014storytelling, beneteau2020parenting, xu2021current, xu2020exploring}. 
Compared to reading and interacting with human partners, children reading with CAs respond to questions with better intelligibility but lower productivity, lexical diversity, and topical relevance \cite{xu2021same}.

Despite consistent findings that demonstrate the potential and benefits of AI technologies to support storytelling practices for children, literature has also shown that AI applications face difficulties in integrating into real-world storytelling or education practices for \review{preschoolers} \cite{su2023artificial}. 
From the technical standpoint, researchers, for instance, have demonstrated existing conversational agents face challenges in generating the expected valid and reliable comprehension story questions and supporting tailored storytelling and interaction \cite{FairytaleQA, yao2022ai, bentley2018understanding, cowan2017can, luger2016like, yang2024talk2care}, which caused children to encounter difficulties in understanding questions from agents, fail to follow the conversation ﬂow, and struggle with appropriate turn-taking when interacting with agents \cite{du2021alexa, pantoja2019voice}. 
In addition, many existing AI-based storytelling tools merely ask questions in a sequential manner from a pre-generated list, which does not prioritize child engagement, as it lacks logical coherence between questions and often fails to provide guidance to children in case of confusion or incorrect answers \cite{Storybuddy, FairytaleQA, yao2022ai}. 
In addition, researchers have also pointed out that existing technologies often exclude the participation of parents \cite{Storybuddy}, who actually have significant roles in practical storytelling \cite{parentrole1, parentrole2}.

The challenges and limitations identified in prior research raise an urgent need to thoroughly understand users' usage and experience of AI-based storytelling technologies, and identify the practical challenges and barriers. In particular, \review{preschoolers} often lack the essential AI literacy \cite{su2023artificial} and skills \cite{resnick2009scratch} to use AI technologies independently.
These insights are, therefore, more critical to inform the development of future AI-based storytelling technologies for \review{preschoolers}, ensuring the alignment with practical storytelling needs and maximizing benefits in AI-based storytelling. Our work joins the ongoing effort to explore the practical challenges and limitations of AI-based storytelling technologies.

\CSCWreview{
Further, as children are increasingly exposed to various AI technologies, researchers have also explored barriers and limitations of AI technologies to children from a broader perspective of child-AI interaction \cite{UNICEF}. Researchers examined the potential social and gender biases of AI technologies to children, and proposed to provide them with opportunities and skills to critically reflect on current AI applications and envision \cite{sharma2023inclusive, zhou2023adapt, wang2024challenges}. Meanwhile, researchers have also examined the impact of increasing digital technologies and the trend of datafication in children \cite{wang2022don, wang2023treat}. They suggested that future AI technologies should be more transparent and autonomy-supportive to children. Given these children-centered findings, researchers proposed to adapt a generic human-centered AI design framework \review{\cite{wang2022informing, zhou2023adapt} }for more ethical, inclusive, and fair AI technology designs. These findings and claims could also be reflected in AI storytelling technologies for \review{preschoolers}. Drawing from the notion of child-centered AI design \cite{UNICEF}, our study deeply examines the usage and impacts of AI-based interactive storytelling technologies, with a particular focus on human (i.e., parents and children) needs, experiences, and perceptions. 
According to our findings, we proposed a series of human-centered design implications for future human-centered, AI-based interactive storytelling technologies. }

\subsection{AI-Based Storytelling Technologies and Parents}

For practical storytelling practices with \review{preschoolers}, parents play the most vital role in, for instance, shaping children's language and comprehension skills development, and strengthening the bond between parents and children \cite{parentrole1, parentrole2}. In AI-based storytelling practices for \review{preschoolers}, parents' roles and functions are even more significant, because \review{preschoolers} often lack the essential literacy \cite{su2023artificial} and skills \cite{resnick2009scratch} to independently use various AI technologies, and they therefore often rely on parents to navigate this AI-based storytelling process \cite{Storybuddy}. According to the study of \citet{lin2021parental}, when integrating AI-based intelligent agents into the story reading and telling scenario, about 63\% interactions are actually driven by parents, and the majority of parents guide children by redirecting their attention or clarifying virtual agents’ prompts.

However, existing AI-based storytelling technologies have been mainly developed and evaluated from children's needs and perspectives \cite{xu2021same, storycoder, EmotionBlock, Wildcard, chubb2022interactive}. In general, attention to parents’ experiences and perceptions of AI-based storytelling is limited. In CSCW and HCI, some prior research has provided formative insight into parents’ objectives for interactive storytelling. \citet{Storybuddy}, for instance, find that parents appreciate the educational potential of new technologies in enhancing parent-child interaction and actively seeking customized content for their children. \citet{boffi2020ding} reports the challenges parents face in educating their children to interact with intelligent storytelling systems correctly and respectfully. Apart from these studies, little prior work has deeply explored parents’ experiences and perceptions of AI-based storytelling technologies. 

Given parents’ significant roles in the adopting and using of AI-based storytelling systems \cite{boffi2020ding, lin2021parental, Storybuddy}, our study provided an empirical study to comprehensively and deeply examine how they perceive and use existing AI-based storytelling tools, their encountered barriers and challenges. Relevant to our study here is \citet{lin2021parental}'s study about parental acceptance of children’s storytelling robots. Through a qualitative study with 18 parents, they demonstrate mixed, though generally positive attitudes of parents towards children's storytelling robots. 
In contrast to their research, our study primarily focused on the challenges our participants experienced and perceived in real-world settings.
Compared to experimental research that occurs in controlled environments and design fiction~\cite{lindley2015back} which uses fictional scenarios and prototypes to explore and envision possible futures for technology, studying the lived experiences of parents in real-world scenarios can provide us with valuable insights into parents’ actual use and perceptions of AI storytelling tools, allowing us to capture the nuances and complexities of how parents integrate such tools into their daily storytelling scenarios, and uncover unforeseen challenges and benefits.
These in-depth, practical insights can help the designers and developers of AI-based storytelling tools to better align with the practical needs and preferences of parents, and understand and address potential issues in fitting AI-based storytelling tools into their storytelling practices. 

In addition, we paid our attention to more general AI-based storytelling technologies, instead of the specific storytelling robots like what \citet{lin2021parental} have done. During our study, we did not specify what kinds of technologies our participants have used, so that the questions could be inclusive of participants to share various levels of experience with different technologies-mediated storytelling practices. Through this broader research scope, we hope to contribute to the general AI-based storytelling technology design by informing designers and developers with in-depth and empirical understandings of the practical challenges and challenges in fitting these technologies into the practical parent-child storytelling scenario and helping them better address these challenges.

\section{METHOD}
This paper aims to gain insights into parents’ practical adoption, usage, experiences, and perceptions toward AI-based intelligent storytelling technologies and tools for \review{preschoolers} between the ages of 3 and 6. 
We are particularly interested in investigating \review{1) current acceptance and usage of AI-based storytelling technologies and tools by parents with preschoolers, 2) their perceived benefits, 3) encountered challenges in using these tools, and 4) their perceptions and expectations of future AI-based intelligent storytelling technologies. }
To this end, we conducted a qualitative interview study with 17 parents. 
\CSCWreview{We iterated through generating codes from collected data and checking and elaborating these codes by collecting more data. We stopped data collection once
all the core variables were saturated.}
The rest of this section details the participants and recruitment, the types of data being collected, and the data analysis process. 

\subsection{Participants and Recruitment}

We looked for participants who 1) were parents with preschoolers between the ages of 3 and 6 and 2) had previous or ongoing experience using AI-based intelligent tools to support storytelling activities. 
\review{We did not specify what AI tools should be in our opinion, and what kinds of AI tools participants had used. We recruited those as long as they considered that they had used storytelling tools with AI features for their storytelling practices with preschoolers.
We also did not specify how much experience they should have, so that the recruitment could be open to people with various levels of usage and engagement in various kinds of AI-based storytelling technologies.
}

The participants were recruited via a snowball sampling approach.
The first author, a native Mandarin speaker, contacted her classmates and friends who had preschoolers. After confirming that s/he had used at least one kind of AI-based storytelling tool, the first author then disclosed our research intention and recruited her/him for our interviews. Seven were contacted, and all agreed to be interviewed. 
To avoid bias from our own social networks, we \CSCWreview{also} sought participants by posting recruitment messages on online forums and communities. We designed the recruitment poster including self-introduction, research intention, recruitment criteria, and contact information of the first author, and team members distributed the poster on their social networks. Twelve reached out directly, and ten of them agreed to be interviewed.


\begin{table}
\centering
\small
\resizebox{0.99\linewidth}{!}{
\begin{tabularx}{\textwidth}{m{0.3cm}m{0.8cm}m{1.8cm}m{1.8cm}m{2.0cm}X}

\toprule[1pt]

\textbf{ID} & \textbf{Roles} & \textbf{Children} & \textbf{Education} & \textbf{Work} & \textbf{Used Tools} \\

\midrule[0.5pt] 
P1 & Mother & \makecell[l]{4-yo son,\\ 10-yo daughter} & Finance, Econ. & Company Staff & \makecell[l]{Interactive Talking Pen (C1), Robot (C5),\\ Little Genius (C3), XiaoDu (C6)} \\

\hdashline[0.5pt/0.5pt]
P2 & Mother & \makecell[l]{6-yo daughter,\\ 12-yo son} & Comp. Sci. & High School Staff & \makecell[l]{HuoHuoTu (C2), iPad (C4),\\ Xiaodu (C6), Tmall Genie (C6)} \\

\hdashline[0.5pt/0.5pt]
P3 & Father & 5-yo daughter & Health Care & Company Employee & \makecell[l]{Talking Pen (C1), HuoHuoTu (C2),\\ Early Learning Device (C3), iPad (C4),\\ XiaoDu (C6)} \\

\hdashline[0.5pt/0.5pt] 
P4 & Father & 5-yo daughter & Business Admin. & Company Manager & \makecell[l]{Jiaojiao (iPad app) (C4), Qiaohu (an\\ educational brand, including TV\\ programs, animated series, interactive \\games, and educational materials)} \\

\hdashline[0.5pt/0.5pt]
P5 & Mother & 6-yo daughter & Automation & CIO & \makecell[l]{Interactive Pen (C1), Interactive\\ Player (C2), iPad (C4), Robot (C5)} \\

\hdashline[0.5pt/0.5pt]
P6 & Mother & 6-yo daughter & Business Admin. & Data Company Employee & \makecell[l]{HuoHuoTu (C2), Smartwatch with\\ story reading apps, iPad (C4)} \\

\hdashline[0.5pt/0.5pt]
P7 & Mother & 5-yo son & Educational Psychology & University Staff & Robot (C5) \\

\hdashline[0.5pt/0.5pt]
P8 & Mother & 6-yo son & Comp. Sci. & Researcher & \makecell[l]{Interactive Talking Pen (C1), \\Seewo (C3), Luka (C5), iPad (C4)} \\

\hdashline[0.5pt/0.5pt]
P9 & Mother & 6-yo son & Finance, Econ. & University Financial Staff & Interactive Talking Pen (C1) \\

\hdashline[0.5pt/0.5pt]
P10 & Mother & 5-yo daughter & Public Admin. & Company Employee & Qiaohu \\

\hdashline[0.5pt/0.5pt]
P11 & Father & 4-yo daughter & Comp. Sci. & University Researcher & \makecell[l]{Interactive Pen (C1),\\ Early Learning Device (C3),\\ Robot (C5), Siri (C6), Qiaohu} \\

\hdashline[0.5pt/0.5pt]
P12 & Father & 6-yo daughter & Education & University Staff & Tmall Genie (C6), XiaoAi (C6), iPad (C4) \\

\hdashline[0.5pt/0.5pt]
P13 & Father & \makecell[l]{3-yo son,\\ 5-yo daughter} & Public Admin. & AI4EDU Institute Staff & MiXiaoTu (C2), iPad (C4) \\

\hdashline[0.5pt/0.5pt]
P14 & Mother & \makecell[l]{3-yo daughter,\\ 7-yo son} & Comp. Sci. & Company Employee & \makecell[l]{HuoHuoTu (C2), Alpha Egg (C5),\\ Tmall Genie (C6), XiaoAi (C6)} \\

\hdashline[0.5pt/0.5pt]
P15 & Father & \makecell[l]{3-yo daughter,\\ 10-yo son} & Psychotechnics & AI4EDU Institute Researcher & \makecell[l]{Interactive Talking Pen (C1),\\ Storyplayer (C2), Alpha Egg (C5),\\ Tmall Genie (C6)} \\

\hdashline[0.5pt/0.5pt]
P16 & Mother & 4-yo son & Lawyer & Company Employee & \makecell[l]{Interactive Talking Pen (C1), Qiaohu,\\ Tmall Genie (C6)} \\

\hdashline[0.5pt/0.5pt]
P17 & Father & \makecell[l]{3-yo daughter,\\ 7-yo daughter} & Education & Kindergartener & Tmall Genie (C6) \\
\bottomrule[1pt]
\end{tabularx}}
\caption{\review{Demographic information of our participants and the tools they have used. (C1-C6 used in the last column corresponded to the categories to which these technologies belonged, with details shown in Table 2.)}}
\label{participant}
\end{table}

Of the 17 interviewees, 10 were mothers and 7 were fathers. 11 of them had one child between the ages of 3 and 6, and 6 had two children, with at least one falling in the age range of 3 to 6. \review{
Their educational and work backgrounds were varied, ranging from information technology, computer science, education, finance, law, and administrative management, to health. 
All of them had used at least one kind of AI-based intelligent tool for storytelling in their opinions. 
Table \ref{participant} shows demographic information of our participants and the tools they have used. 
}

\subsection{Data Collection}

The interviews were semi-structured, including questions about 1) their daily storytelling practices and experiences, 2) the AI-based intelligent storytelling tools they have used and how they experienced these tools 
\CSCWreview{(when the technologies and tools reported by participants did not align well with research communities' definition of AI technologies, we explored as much as possible the ``intelligent functions'' that parents considered from these technologies and tools) and prompted them by illustrating typical ``AI technologies''}, 3) the benefits and challenges they experienced, and 4) their expectations of future intelligent tools. 
For parents who have used LLM-based tools (e.g. ChatGPT), \review{we specifically asked 5) how they considered if LLM-based technologies were used to support preschooler-oriented storytelling. }
All interview questions were designed to be general so that the questions could be inclusive to participants to share various levels of experiences with different technology-mediated storytelling practices.

The interviews were mainly conducted in Mandarin by the first author. 12 were conducted via WeChat Voice call and Tencent Meeting, and 5 were conducted in person. Each interview lasted for 40-60 minutes, with a 100 CNY honorarium. 
During the interview, we encouraged participants to provide detailed demonstrations of how these tools functioned and how they used them. When some interesting points or prior experiences were mentioned, we asked follow-up questions for more details and concrete examples.
With participants’ permission, we also obtained other forms of data to triangulate their interview responses. These data included 1) screenshots of mobile applications the participants have used and 2) related photos or videos they took, if any.

\subsection{Data Analysis}
With the participants’ permission, we recorded the audio of all interviews and transcribed them into Chinese verbatim after the interviews for analysis. We followed thematic analysis as outlined by \citet{clarke2015thematic} to analyze the main themes in the interview data in an inductive way \cite{thomas2006general}. Three authors participated in the data analysis process. We started analysis while the data was being collected. During the open coding phase, three authors independently reviewed the collected data and generated codes pertaining to our research questions. We had regular meetings on a weekly basis to discuss codes and ensure reliability. After this stage, we generated the initial code list, capturing participants’ practices, motivations, and perceptions related to their daily storytelling actions; their usage, motivations, and experiences related to the intelligent storytelling tools they had used, as well as their perceptions and expectations to the general AI-based storytelling tools. 

Based on this code list, we then re-focused our analysis at the broader level of themes, and synthesized and grouped the codes to create higher-level overarching themes. The initial analysis centered on themes related to 
\review{storytelling tools parents had used}, parents' adoption, usage and experiences of these tools, the difficulties and challenges they have encountered, their children's adoption, usage and experiences of these tools, parents' perceptions and concerns about LLM-based AI tools, and parents' expectations of future AI technologies. 

\CSCWreview{As analysis proceeded, we found that, due to our open criteria to ``what AI tools should be and what kinds of AI tools participants had used'', some tools reported by our participants, such as Huohuotu, story player, etc., were not aligned well with research communities' typical definition of AI technologies (referring to technologies that enable computers and machines to learn, read, write, create and analyze, simulating human intelligence and problem-solving capabilities\footnote{https://www.ibm.com/topics/artificial\-intelligence}
) on the one hand, and some, such as part of applications in iPad, were not used for story reading and telling, but for learning on the other hand. 
Considering the research aim of ``\textit{gaining insights into parents’ practical adoption, use, experiences, and perceptions toward AI-based storytelling technologies and tools, through which to shed light on how we can better design future AI tools to support parent-preschooler interactive storytelling}'', we removed codes regarding parents’ practical usage code of these tools to focus the analysis more on the codes relevant to our research questions. We still retained data on participants' experiences with these technologies because we considered that these data expressed parents' actual needs for storytelling technologies.}
In addition, we realized that the perceptions of using existing intelligent storytelling tools and LLM-based AI tools were closely related. We then further focused on themes that reflect similarities and differences between existing intelligent storytelling tools and LLM-based AI tools.

Through several rounds of analysis and discussions, three authors reached an agreement and generated our final satisfactory thematic map of the data, which included four primary themes: 1) parents' attitudes and perception of AI-based storytelling with preschoolers, 2) their perceived benefits, and encountered 3) interactive and 4) algorithmic challenges in practice. 
 In the following section, we present the full details of these themes. Throughout, we use representative quotes from our participants, translated from Chinese into English by the first author, to illustrate our points. To protect our participants’ identities, we use P1, P2, etc., to denote our study participants.

\subsection{\review{Positionality of the Authors}}
\review{We acknowledge that, for qualitative research, the research perspectives and approaches are shaped by our own experiences and positionality~\cite{pascoe2022reflections}. Specifically, we are researchers living and working in the China and U.S., primarily researching human-centered AI and the perceived trustworthiness, fairness and effectiveness of AI in education.
We come from a mix of disciplinary backgrounds, including Computer Science, Learning Sciences and Technology, Education, and HCI, which we have drawn on to conduct prior research into sociotechnical approaches to human-centered AI design practices.
We have experience working with school teachers and district personnel on technology integration, and AI practitioners on projects related to human-centered educational technology design. 
In general, we recognize the tremendous potential of the rapidly developed AI in promoting societal progress and enhancing people's quality of life and work. However, we disagree with the notion that AI will replace human roles, such as the role of parents in children's storytelling. We firmly believe that the optimal design approach should be human-centered, offering greater flexibility and adaptability. Our results confirm the validity of our hypothesis.}

\subsection{Ethical Considerations}

\CSCWreview{All researchers received formal ethics training, and were granted official certificates from the ethics committee in their institute.
During the study, we took careful steps to protect user privacy. Before we started interviews, we informed participants of our intention and background information, and got permission from them. All data collected during our study was used in an anonymized way, i.e., there was no link between the collected data and an individual user. We strictly abide by the confidentiality agreement, which promises the collected data is only used for this research and any form of information leakage is not permitted.}

\section{Findings}

\begin{table}[]
    \centering
    \begin{tabular}{c}
         \includegraphics[width=1.0\linewidth]{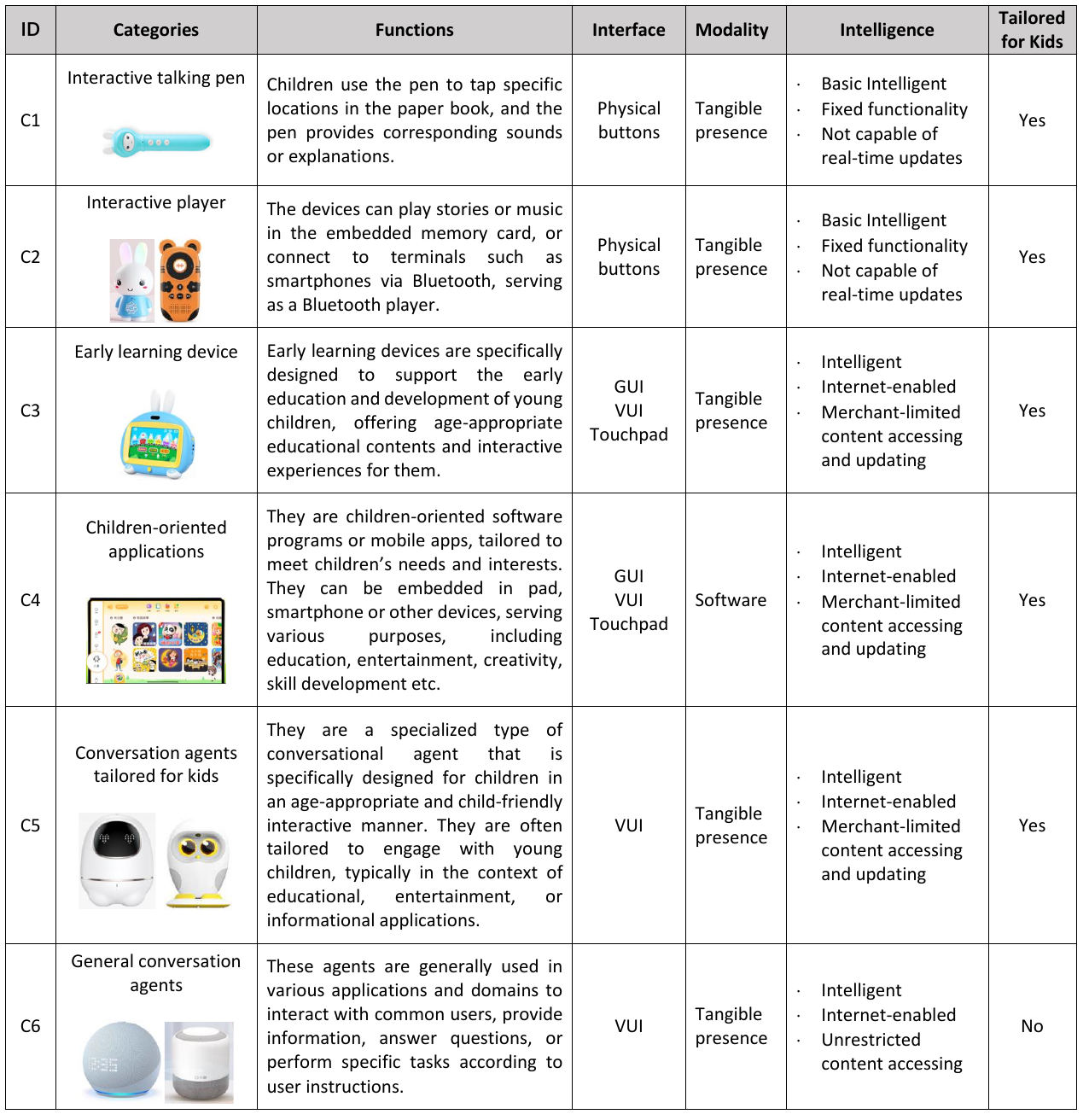} 
    \end{tabular}
    \caption{Some AI-based storytelling technologies reported by our participants}
    \label{tab:tool}
\end{table}

Our findings suggested that it had been very common among \CSCWreview{our participants to utilize various types of AI-based technologies and tools} to facilitate their daily parent-children storytelling activities (as shown in the right column of Table \ref{participant}). Among 17 parents we interviewed, 14 had used more than two types of technologies to support their daily parent-children storytelling practices.
According to the interaction modes and functionalities, we roughly classified these tools into six categories. Specific details can be found in Table \ref{tab:tool}. Some of these tools were aligned well with research communities' typical definition of AI technologies, while some others, such as tools of C1 and C2, were not. 
In the following, we will elaborate on the experiences, perceived benefits, and encountered challenges of our participants when using these tools.

\subsection{\CSCWreview{Parents’ Attitudes and Expectations to AI-Based Interactive Storytelling}}


\subsubsection{Parents’ Attitudes to AI-based Storytelling Tools: Cautiously Positive}
Generally, our participants expressed a cautiously positive attitude toward existing AI-based storytelling tools. That is, although they believed AI's potential in supporting interactive storytelling for preschoolers and reported some benefits of existing tools they perceived (see section \ref{finding:value} for details), they emphasized a series of concerns to and limitations of these tools (see section \ref{finding:challenge1-interactive} and \ref{finding:challenge1-algorithm} for details). 

\CSCWreview{Further, none of the participants believed AI technologies, no matter how they became, could replace the role of parents in interactive storytelling scenarios, because ``\textit{it cannot support interactions with emotion and warmth, and cannot replace parents’ love, care and hugs}'' }[P13, a father with a 3-year-old son and a 5-year-old daughter]. They emphasized that \CSCWreview{interactive storytelling with preschoolers was a very child-centered comprehensive process. Through this process, they hoped children could perceive the world, learn different knowledge and skills, and feel different cultural contexts, and hoped to guide children's active thinking, cultivate their reading habits, or \textit{just be there and spend parent-child time with my child} [P5].} Yet, AI-based tools \review{only paid attention to reading the story or teaching knowledge}. 
P9, a mother with a 6-year-old son strongly opposed the existing trend of using AI technologies \review{to read stories for preschoolers}. She believed that children's growth should be accompanied by parents, rather than \review{``\textit{a smart machine}''}.


\begin{quote}
    \textit{``You propose to design a super smart machine and ask me to use it to educate my son. Why? I don't understand. I think the most important thing is the interaction between parent and child, between human and human. Mothers should not find a machine to accompany their children, but rather `I accompany you'. A machine is just a machine, no matter how advanced it is, it is still just a machine.''}[P9]
\end{quote}

In addition, many of our participants thought that \CSCWreview{AI-based interactive storytelling} was not real storytelling and reading. They emphasized that storytelling and reading in the real world was not just about reading stories, asking questions, or teaching some knowledge, but also the reading habits, thinking methods, and even postures and attitudes of reading. Parents need to pay attention to all of these aspects, but AI tools only read the stories and teach knowledge.

Significantly, our findings suggested that parents' attitudes were somewhat correlated with their educational background, work experience, and parenting philosophies. 
Specifically, we found \CSCWreview{parents' educational and occupational backgrounds influenced their adoption and attitudes to AI-based storytelling tools to some extent}. 
Among our participants, five were engaged in technology-related professions, or with technology-related educational backgrounds (P5 was a CIO of a company, P6 worked in the field of digital healthcare, P8 and P11 were involved in computational science research, and P13 held a managerial position in an intelligent education research institution). In contrast to the other participants, these five participants exhibited more positive attitudes and higher expectations for \CSCWreview{AI-based interactive storytelling technologies}. They believed that despite the current limitations of AI technologies, there was great potential for future technologies to address these limitations and better assist parent-children \CSCWreview{interactive} storytelling. They expressed their eager anticipation of the emergence of such technologies. In contrast, four participants with educational backgrounds or occupations in the educational field (P2, P7, P12, and P17), hold relatively conservative attitudes toward AI-based \CSCWreview{interactive} storytelling tools. During the interviews with them, they explicitly expressed strong dislike and even disapproval of preschoolers' use of AI-based tools, with the reasons of \CSCWreview{tentions between preschoolers' limited discriminatory abilities and uncontrollable AI (P2), between preschoolers' cognitive development level and AI's virtual interaction modes (P7), between preschooler' free, personalized development and AI's overly direct goal orientation and standardization (P12), etc.}

\review{Parenting philosophies also significantly impact \CSCWreview{our participants'} attitudes toward AI-based storytelling technologies. Throughout our interviews, participants often articulated their viewpoints using statements like ``\textit{I believe that education for children should be XX; in this philosophy, AI should XX.}'' Our data indicated that parenting philosophies had a more direct impact on parents' acceptance and use of AI tools than educational or occupational backgrounds. For example, despite P2 having a background in computer science and P15 working in an intelligent educational research institution, their attitudes toward AI tools were not particularly positive. Instead, they lean towards the idea of parents personally accompanying their children in storytelling, rather than relying on a smart machine.}


\subsubsection{Parents’ Expectations to AI-based Storytelling Tools: ``\textit{Being Parents’ Secret Weapon}''}

While all our participants agreed that AI technologies absolutely couldn't replace parents' role in practical interactive storytelling scenarios, they expected AI-based tools to be their `secret weapon' to help free up their physical labor, assist in meaningful parent-child interaction, and help better understand children and find resources according to their personalized characteristics and needs.

\textbf{\textit{Freeing up physical labor}}. Many of our participants expressed that telling stories to children was a very tiring job, and it became even more difficult if interaction was required. They therefore expected a machine could help them read stories, then they could focus on more valuable interactions with children. P14, a mother of a 7-year-old son and a 3-year-old daughter, expressed her strong expectation of technology-supported story reading tools: 
``\textit{Reading for 20 minutes every day often makes me exhausted. It is a kind of physical exhaustion of the mouth. Interactive storytelling is more tiring. I'm often tired when I come home from work. Sometimes, when I tell my child a story, I end up falling asleep myself. If a robot can read the story, it will be very great. } ''

\textbf{\textit{Assisting meaningful parent-child interaction}.} Given the practical challenges many parents faced in conducting meaningful and high-quality interactions with their children, our participants hoped AI tools could \review{assist} their parent-child interaction, such as generating valuable interactive content, and inspiring parents on how to interact with children. As P9 said, ``\textit{If my son asked me a question and I didn't know the answer, I would tell him, `Wait a moment, let me ask my secret weapon'. After finding the answer, I would tell him. I thought this was a very cool thing.}'' She emphasized that the important part was that the secret weapon told the answer to parents, and helped parents to better interact with children, rather than the machine directly interacting with children.

\textbf{\textit{Helping to better understand children and find resources according to their personalized characteristics and needs}}. With the explosive growth of digital learning resources, providing valuable and suitable learning materials for children at specific stages was a common challenge faced by parents. They consequently hoped AI technologies could help them better understand their children and find valuable and suitable resources according to their personalized characteristics and needs. Many parents in our study expressed that they were not satisfied with many currently used tools, but they still used them mainly because ``\textit{I don't have to find resources by myself}.'' 

In summary, our participants expressed varied attitudes toward AI-based storytelling but shared similar expectations of it being parents' assistant. Then, do the existing AI-based storytelling tools meet parents' expectations? How effective are they in practical use? What is the parents' experience with their usage? We now discuss these questions in more detail in the following sections.




\subsection{Perceived Value of Using AI-Based \CSCWreview{Interactive} Storytelling Technologies}
\label{finding:value}

When reflecting on the experience of using AI-based \CSCWreview{interactive} storytelling tools to associate daily parent-child storytelling, our participants talked about some benefits of these technologies they perceived.
\CSCWreview{For example, while all our participants agreed that interactive storytelling was a very important and beneficial parent-child activity for preschoolers, most of them expressed this conversation-rich interactive storytelling was full of practical challenges.} As P9, a mother with a 6-year-old son, said, ``\textit{I really want to interact with my son, but I don't know what story content should be interacted. I even can't understand the questions he asks me.}'' Some participants also noted that engaging in daily, high-quality interactive storytelling, as they expected, represented significant challenges in terms of time, energy, and physical stamina. This was particularly true for today's young parents who were often immersed in fast-paced and demanding work. 

 \review{Under such a situation}, AI-based storytelling tools, \review{worked} as more captivating and technologically advanced toys, could help alleviate these challenges to some extent. Our participants expressed that although existing AI-based technologies they have used didn't fully meet their expectations in supporting children-oriented interactive storytelling (see more details in the following subsections), they did find them helpful in interacting with children when they were occupied (i.e., playing with children or telling stories).
 Meanwhile, these technologies could partially relieve parents from those repetitive and monotonous story-reading tasks, allowing them more time to engage in the desired high-quality interactions. As P6, a mother of a 6-year-old daughter, said, 

\begin{quote}
    ``\textit{I wish I could spend an hour every day telling her stories, but I don't have enough time and energy. I can't always be present with my child. So, how can my child spend that time? That's where XiaoAi (a smart speaker similar to C6 in Table~\ref{tab:tool}) comes in. At least it can read stories to her, or engage in conversations. Although it doesn’t work very well at the moment, it allows me to dedicate my efforts to more challenging tasks, such as interacting with my child and responding to her questions.   
    ''} [P6]
\end{quote}

At the same time, many of our participants stated that compared to plush toys and the like, these AI-based tools were relatively ``better'' toys for children, because they could bring ``\textit{novelties and a sense of technology}'' [P8, P13] to children, which parents couldn’t bring. \review{Most existing storytelling tools had some children-tailored interactive designs, such as immersive storytelling blending multiple modalities (``\textit{Children could touch, hear, see, and talk with it.}'' [P6]), and vivid presentations (``\textit{The sound of the storyteller is very pleasant and full of emotions and varied intonations, and the animals can move and make sounds}'' [P1]), which worked well to attract children to participate in the interaction, improving their storytelling experience significantly. }

In addition, most of these technologies integrated more or less learning resources, or had some child-tailored interactive designs, which allowed children to learn something or develop certain abilities while playing. As P8, a mother of 6-year-old son, said: 
``\textit{You always need to buy something for the child, right? Do you buy a bunch of Ultraman or cloth dolls, or these intelligent things? No matter how the practical effect is, it provides the possible opportunities to cultivate the creative thinking abilities of children, or learn something when playing.}'' [P8]

Despite these perceived benefits, our participants considered that 
most of the existing storytelling tools they had bought and used were hard to consistently engage preschoolers and often struggled to achieve the intelligent features claimed by the products. The practical effectiveness and efficiency in supporting quality storytelling were limited.
 Our participants reported several challenges with these tools that hindered their practical effectiveness. We generally categorized these challenges into two types - interactive and algorithmic, detailed in the following two sections.

\subsection{Interactive Challenges in Supporting Storytelling Aligned with Preschoolers' Cognitive Characteristics}
\label{finding:challenge1-interactive}
\CSCWreview{Interactive challenges mainly refer to challenges and dilemmas caused by interactive ways and interfaces existing AI-based storytelling tools used.
Our participants stated that existing AI-based storytelling tools they had used didn't consider preschool-age children's cognitive characteristics and their abilities to accept and use these technologies, which hindered the tools' practical effectiveness. }

\subsubsection{Preschoolers are Difficult to Use AI Tools Proactively, Independently, and Consistently}

According to our participants, preschoolers often exhibited short attention spans, lacked self-learning and self-management abilities, etc. These characteristics made it hard for them to use AI-based tools actively, independently, and consistently. Most of our participants thought it was unrealistic to expect preschoolers to use these tools to read a storybook independently. 
Instead, most tools require parent-engaged usage processes. For example, the use of an interactive talking pen (C1 in Table~\ref{tab:tool}) required parents to ``\textit{take out the pen, turn it on, find the corresponding book, and flip to the specific story} [P8]'' and then the children could use it to read a story. ``\textit{Without parents’ operations and guidance, \review{preschoolers} are unlikely to proactively use these tools to read the story}'' [P8]. 
P15, who had purchased two conversational agents tailored for kids (Similar to C5 in Table~\ref{tab:tool}) for his 3-year-old daughter and 10-year-old son, also expressed that "\textit{today’s children have too many interesting things to do. Without parents' engagement, they may only use them (conversational agents) for a short time out of curiosity and then turn to do other things} [P15]." 

Further, preschoolers' short attention spans made it challenging to consistently and sustainably use any tools with focused attention. Whether and how parents were involved affected the practical effectiveness of tools. Many parents in our study expressed that there was a notable difference in effectiveness between children using these tools independently and when accompanied by parents. 
P5, a mother of a 6-year-old daughter, considered that, ``\textit{if parents are not involved, even for the best product, it is difficult to sustain children’s interest and attention, and motivate them to read a story with it for 15-20 minutes steadily}''. P8, who had purchased many AI-based storytelling tools for her son, shared her experience with these tools.

\begin{quote}
``\textit{Our (parents’) involvement or not has a completely different using effect. I can let these tools tell stories. Tools, like Luca (C5 in Table~\ref{tab:tool}) and Xivo (similar to C4 in Table~\ref{tab:tool}), are well-designed for this task. Objectively, they perform well in terms of design, interaction, and storytelling. The issue is, preschoolers have short attention spans, without any active learning consciousness. It is their basic characteristic. ... 
I've heard that some children start to slack off once they've mastered the tool. They can quickly complete the reading task by repeatedly clicking the progress bar.}'' [P8]
\end{quote}

\subsubsection{Preschoolers Need the Sense of Authenticity and Constructiveness}

The virtual interactive modality, without the sense of authenticity and constructiveness, was also reported by our participants as one major reason that current AI-based storytelling technologies couldn't sustainably attract children's attention and achieve the expected storytelling effect. They expressed that preschoolers were still in the stage of development of motor thinking, and their abstract thinking ability was not well developed. This thinking development stage determined that they needed more physical and real interactive processes with a sense of constructiveness, instead of the virtual way, as most existing intelligent tools used. 
P15 used his son's experience of playing with building blocks to explain this sense of authenticity and constructiveness: 

\begin{quote}

``\textit{My son can play with building blocks for more than two hours. He has a sense of reality that he is really building something. Whatever he thinks in his mind, he can immediately construct it, such as a castle, a kitchen, or a gun} [P7]''.
\end{quote}

Meanwhile, compared to human-human interaction, our participants also recognized that virtual interaction methods lacked the necessary emotional connection, which resulted in intelligent agents struggling to effectively sustain children's attention and engagement. As P15 said,

\begin{quote}
    \textit{In the beginning, there was a sense of novelty. My child interacted with them (C5 in Table~\ref{tab:tool}) actively, saying things like `Egg, come with me; Egg, sing me a song; Egg, tell me a story.' But he only played with them for three days. Currently, they are just expensive toys, sitting in the corner. Children can't establish an emotional connection with them. I think things that can be touched and felt will provide more satisfying experiences for children.} [P15]
\end{quote}

Additionally, our participants also considered systematic and contextual information as significant in constructing children's cognition of the stories, but lacked in AI-based storytelling scenarios. 
They stated that in traditional paper books-based storytelling, the storybooks often possessed coherent story structures and contextual situations. Throughout the reading process, children could perceive this contextual information and construct the stories in their minds through flipping the pages back and forth. However, AI agents only presented the story plot at a specific moment, devoid of any contextual information. This limited their ability to develop children's comprehensive understanding of the story, subsequently impacting their engagement and interests.

\subsubsection{Preschoolers Need Interaction, Instead of a Correct Answer}

Interaction and stimulating children’s active thinking were reported by our participants as one primary expectation of \CSCWreview{interactive storytelling}, even more important than learning knowledge. Through \CSCWreview{interactive} storytelling, they desired children could actively participate in the process of imagination, emotional expression, and critical thinking, which were helpful to enhance their creativity, empathy, and problem-solving skills. P13, for instance, showed us an \CSCWreview{interactive storytelling} process he conducted with his 5-year-old daughter, which illustrated this interacting process. 

\begin{quote}
    ``\textit{One day, I told my daughter the story `Monkey move corn', in which the monkey initially held the corn. After seeing a peach, the monkey held the peach and threw the corn away. After seeing a watermelon, it then held the watermelon and threw the peach away. ...
    After seeing a rabbit, it chased the rabbit and threw the watermelon away again. 
    In the end, the monkey didn’t catch anything. This classic story has many inherent truths, such as the importance of having a clear goal and being focused on it. However, I didn’t tell my daughter this moral directly. Instead, I asked her, `if you were the monkey, what would you do?' To my surprise, my daughter responded with great insight. She said, `if I were the monkey, I might have three ways to solve the problem. First, I would bring a bigger bag to take everything. Second, I would pick corn today, peaches tomorrow, and watermelons the day after tomorrow. Third, I would call my friends to come with me, then we could gather corn, peaches, and watermelons together.'} [P13] ''
\end{quote}

Based on this story, P13 and many other participants emphasized that meaningful interactions should aim to stimulate children's active thinking, rather than simply imparting a fixed moral or lesson. 
\CSCWreview{However, most current AI-based storytelling tools pay primary attention to supporting correct question-answering, i.e., providing a `correct answer' to children's questions. They were completely unable to support their desired interactive storytelling. For this reason, some of our participants strongly rejected the tools with explicit educational purposes, because they might limit children’s thinking and self-reflection.}





\begin{quote}
   ``\textit{Pre-school storytelling is not mainly for learning knowledge, but many existing tools focus on `teaching you something'. They are very eager to impart knowledge to children. I don't like them at all. A story is just a story. You just need to tell a story. 
   What children can learn or think from the story depends on the interactor and themselves.
   }''[P12]
\end{quote}

\CSCWreview{Given these interactive challenges, our participants suggested that AI-based interactive storytelling technologies should carefully consider preschoolers’s inherent characteristics and cognitive development level. }Meanwhile, technologies should guide children’s active thinking, instead of just telling them the correct answer or specific knowledge.

\subsection{Algorithmic Challenges in Supporting Child-Centered, Intelligent, Personalized, and Heuristic Storytelling}
\label{finding:challenge1-algorithm}
Algorithmic challenges 
primarily arose from the limited ability of AI algorithms to support child-centered, heuristic, personalized, and adaptive storytelling practices, which further caused the limited usability and effectiveness of existing AI-empowered tools in facilitating practical, parent-expected intelligent storytelling for preschoolers. 

\subsubsection{Challenges in Supporting `Intelligent' Interaction}

Silly and limited-intelligent interaction was one major algorithmic challenge our participants experienced when they used AI-based tools, no matter children-tailored tools (e.g. C3, C4, C5 in Table~\ref{tab:tool}) or general intelligent tools (e.g C6 in Table~\ref{tab:tool}). For children-tailored tools, our participants commonly considered their intelligence was limited, and interaction ways and content were notably inflexible, adhering to a fixed structure and only permitting content access and updates within the limitations set by the merchant. The experience of P1 using Little Genius (similar to C4 in Table~\ref{tab:tool}) illustrated this limited intelligence very well: ``\textit{its content is almost static, and the interaction remains unchanged. Whether you open it this year or next year, it remains the same. It lacks adaptability and fails to adjust according to the child's developmental needs.}''

For general intelligent conversational agents based on massive corpora, such as C6 in Table~\ref{tab:tool}, our participants considered that, although they performed commands well, their abilities in supporting children-oriented dialogue and interaction were still very limited. P6’s experiences of using XiaoDu (C6 in Table~\ref{tab:tool}) illustrated this limited intelligence very well:

\begin{quote}

``\textit{Whether it's XiaoDu or XiaoAi, its intelligence level is not that high. It often struggles to understand a child’s command and falls short in delivering appropriate responses, especially when the child's speech is unclear. ...
For instance, when you ask it to narrate a story about Journey to the West, it might veer off-topic and present various unrelated products related to Journey to the West.}'' [P6]

\end{quote}

In particular, our participants highlighted that in real-life scenarios, children's interests and questions were frequently diverse. Yet, current AI-based storytelling technologies were more like search engines. They could effectively answer questions that could be queried through search engines, but failed to answer those that couldn't be queried. \CSCWreview{As P8 evaluated the function of Alpha Egg (C5 in Table~\ref{tab:tool}): 
``\textit{It's just a search engine. When you ask who is Osa, it searches online and reads the answer to you. Once your question can’t be searched online, then it fails to answer.}'' [P8]}
Even with more intelligent generative AI technologies, parents in our study still expressed their concerns about their capacity to genuinely grasp and decipher preschoolers's intentions and furnish responses. They expressed \CSCWreview{concerns} about whether and how generative AI technologies could answer children's endless and personalized ``why'' questions well, because many of these questions actually didn't have standard answers. As P6 said,

\begin{quote}
    ``\textit{Child's questions are often very strange. They may ask why witches are female, why witches have pointy noses, why witches are thin instead of fat, why Harry Potter is named Harry Potter, why he rides a broomstick instead of a vacuum cleaner... There are no standard answers to these questions. How can machines answer these questions? 
    }'' [P6]
\end{quote}

\subsubsection{Challenges in Supporting Children-Appropriated Interaction and Content Generation}
Whether AI tools could support children-appropriated interaction or generate children-appropriated content was another significant algorithm challenge reported by our participants. 
Our participants indicated that when they used tools such as C5 and C6 in Table~\ref{tab:tool} to answer children's story-related questions, the generated answers were often serious and professional, not specifically tailored to children’s cognitive level and psychological age, which caused it difficult for Children to understand. P15’s experience of using XiaoAi to answer children's questions illustrated this very well: 

\begin{quote}
``\textit{My child asked me what `shame' was in the story, and I asked XiaoAi. It gave an explanation. But it was like a dictionary, just reading out the explanation in the dictionary. If you ask it what a water tank is, it can also give a very scientific answer. But such kind of answers were too serious for the young child to understand.}'' [P15]
\end{quote}

A similar concern was also reported by participants who had used ChatGPT to answer children’s questions. They expressed that the answer generated by ChatGPT was not very suitable for \review{preschoolers}, particularly in terms of knowledge level, language structure, and presentation. As P8 said, ``\textit{The content is often too long and professional. Many terms are unfamiliar to \review{preschoolers}. If it was used to interact with children, it needed to be shorter and simpler. I think this is a big technological challenge. There are thousands of questions, and how can it answer them in the concise and appropriate language that children can understand?}''

\subsubsection{Challenges in Supporting Personalized Interaction}

Personalized interaction was reported by our participants as the basic interaction requirement of AI-based storytelling technologies. 
They emphasized that, in practice, each child's characteristics, interests and questions were different, ``\textit{some children are receptive, while others are more proactive}'' [P10], and ``\textit{some children prefer to read by themselves, while others like you (the parents) to read for and interact with them}'' [P5]. These personalized features caused, even if children saw the same story, their points of interest might be completely different. ``\textit{With the same story, different children may focus on different things. Boys and girls think differently, my child and her tablemate in kindergarten also think differently}'' [P15]. 
\CSCWreview{Further, our participants emphasized that one child's character, interests and focuses also dynamically change. Many participants told us that their children read the same story repeatedly, but their understanding and points of interest about the story content changed continuously as they grew older. }

However, although many existing AI-based storytelling technologies
claimed that they supported personalized interaction and learning, in our participants' view, current personalized interaction was very coarse-grained (e.g., ``\textit{first graders may read these books, second graders may read these books}'' [P12]), and couldn't fully support their expected personalized interaction. They emphasized that true personalization for \review{preschoolers} needed to consider various factors such as children's real-time interests, needs, characteristics, personalities, etc., not just the grade and age. Meanwhile, these technologies didn't provide the functions of automatically perceiving personalized characteristics of children. Instead, they often relied on parents manually inputting some basic characteristics of the children and based on these characteristics to conduct ``\textit{coarse-grained personalization}''. This made it more difficult for the personalized features to become a reality \CSCWreview{in practical interactive storytelling scenarios.}
 
\CSCWreview{Even some technologies employed recommendation algorithms to support personalized story and content recommendations, our participants considered them as ``\textit{almost incapable of providing human-centered personalized recommendations}'' [P8]. They thought these tools worked more like ``\textit{the child reads a story today, then the system recommends the similar stories for the next few days}'' [P4], which might ``\textit{greatly limit the personalized development of \review{preschoolers}}'' [P4]. } 


\subsubsection{Challenges in Supporting Situated and Adaptive Interaction}

\CSCWreview{According to our participants, the practical storytelling was a dynamic and situated process, in which parents had a dynamic awareness of children’s real-time interests and attention. Based on children's feedback, parents dynamically adjusted the interaction content and ways. As P9 said,}


\begin{quote}
``\textit{We didn't know what our child liked at the beginning. We adjust the interaction based on his feedback and state. For example, from his eyes and body language, we can learn that he doesn’t want to read anymore, he wants to go out to play, or he is not interested in this story. If we ask a question and he doesn't react, we will change the question or the wording. Through this process, we gradually know what content and interactive methods he likes.}'' [P9]
\end{quote}

Nonetheless, our participants stated that the majority of existing AI-based storytelling technologies were unable to perceive children's real-time situations very well, which impacted the actual effectiveness of these tools directly. ``\textit{Machines have a lot of knowledge and just want to impart it to children. They have no idea whether children want to learn or are interested in what they are saying.}'', as P12 said. 


\subsubsection{Challenges in Supporting Children’s Active and Diversified Thinking}

Throughout our study, our participants consistently emphasized that every child was unique. However, current technologies, they considered, fell short of adequately supporting the diverse development of children. For children-tailored conversational agents (e.g. C4 and C5 in Table~\ref{tab:tool}), which were built on specialized children's corpora and could ask children questions proactively, the interaction was primarily based on relatively fixed patterns and language models, which evaluated by our participants as ``\textit{limiting children's active thinking and diversified development}'' [P11]. 
They thought children should engage actively in thinking and posing questions based on their curiosity, rather than relying on a machine to ask pre-determined questions and provide predefined answers. P4 explained this opinion through his using experience to C4 in Table~\ref{tab:tool}: 

\begin{quote}
``\textit{Each story with it is followed by a universal truth, such as the stepmother is evil. But the child may be interested in stepmother's beautiful dress, and its color. The process of proactive thinking and questioning is valuable. Even though many of the child's questions appear strange, silly, or stupid, they are actually their own questions and expressions of their interests.}'' [P4]
\end{quote}

For general intelligent conversational agents based on massive corpora, such as C6 in Table~\ref{tab:tool}, and even more intelligent generative AI technologies such as ChatGPT, many of our participants, such as P4 and P7, still expressed their rejection of using them, with concerns of ``standard answers'' limiting the diverse development of children. They thought that, no matter how intelligent, AI technologies were fundamentally built upon extensive data-driven pattern recognition, i.e. mining rules based on massive data and providing `standardized' and `correct' answers, which were definitely not suitable for preschoolers, and even potentially mold them into fixed paradigms. 

\begin{quote}

``\textit{I don't support using ChatGPT to tell stories or interact with children. ChatGPT can answer many questions, hundreds of thousands of questions, but these questions and their answers have already been summarized by predecessors. It standardizes these things and gives them to children, which actually limits children's diversified thinking and development.}'' [P4] 
\end{quote}

\subsubsection{Challenges in Supporting Authenticity, Civilized, Safe and Ethical Interaction}

Challenges in supporting authenticity, civilized, safe, and ethical interaction primarily pertained to general AI technologies such as C6 in Table~\ref{tab:tool}, and the booming generative AI technologies due to their lack of transparency and insufficient controllability \cite{wu2022ai}. Our participants expressed that preschoolers did not have the ability to discern multiple information. Given the insufficient controllability, they were worried about what kind of conversations children would have with such a smart robot, and who could ensure the authenticity of the output content.
 
The value orientation of the content output by ChatGPT was also another concern expressed by our participants. P2 was a Christian and explained this ``value orientation'' concern to us: ``\textit{in our family, we prioritize this aspect as we have strong beliefs. We focus on concepts that align with our faith, and scientific knowledge that contradicts our beliefs is not widely accepted. For example, we adhere to the creation theory rather than the evolution theory. I am unsure whether ChatGPT has the capability to accommodate or respect these specific preferences}''[P2].

In addition, \CSCWreview{parents in our study} were also concerned about LLM-based AI technologies' ability to recognize and guide children correctly when they asked inappropriate questions. As P2 said, ``\textit{if a child asks how to not do homework or how to beat my classmate, how would the machine respond, and can it allow the children to ask that question?}'' [P2]. The civility of language was also a concern for parents. Given children's strong learning and imitation abilities, our participants worried if ChatGPT used uncivilized or inappropriate language styles, it would have a very negative impact on children. Based on these concerns, many of our parents stated that they currently did not allow their children to access uncertain content on the Internet without their control. ``\textit{Nowadays, many things seem uncontrollable, and the world is very dangerous. I will never let my children face the evil world without my control}'' [P2].


\section{Discussion}
This paper has illustrated \CSCWreview{how parents with preschoolers in our study} experience and perceive existing AI-based storytelling technologies and tools in practice. Our findings indicated that, despite some perceived value, existing AI-based storytelling tools our participants had bought faced numerous practical challenges in supporting their expected \CSCWreview{interactive storytelling}. We categorize these challenges into interactive and algorithmic and elaborate on them. According to these findings, we now delve into parents' expectations for ideal parent-child interactive storytelling and propose research implications for future AI-based storytelling research.



\subsection{\CSCWreview{Re-Conceptualizing Human-AI Interaction Paradigm in AI-based Storytelling Scenarios: Taking Parents as ``Beneficiary''}}

\CSCWreview{Most existing AI-based storytelling systems \cite{UNICEF} primarily focus on child-AI interaction. However,
our research has revealed several challenges with this child-AI interaction paradigm.} Specifically, preschoolers typically lack the suitability to use tools for storytelling independently and consistently. Meanwhile, the capabilities of existing AI tools are insufficient to meet parents' expected storytelling. Furthermore, the process of storytelling for preschoolers is inherently a parent-children interactive experience, with one primary goal of enhancing parent-child relationships, as reported by our participants as well as existing literature \cite{wright, chang2016mother, peck1989using}. However, this aspect is not well-supported under the existing child-AI interaction paradigm.

\CSCWreview{In CSCW and related fields, researchers are increasingly acknowledging the limitations of this child-AI interaction paradigm \cite{Storybuddy, SignBright}, and also emphasizing the positive impact of parental involvement in children's use of AI for storytelling \cite{cagiltay2022understanding, luka}. Our study, echoing this literature, further emphasizes the importance of recognizing parents as direct beneficiaries of AI tools, and supporting an interactive paradigm among parents, children, and AI tools, rather than only focusing on child-AI interaction.}
Based on our findings, we specifically propose the following design implications.

\textbf{\textit{Taking Parents as Beneficiaries: Support, not Substitute.}} We argue that the role of AI tools should be assisting parents in storytelling, rather than substituting their role entirely.
Despite the limitations of existing AI tools, our participants expressed expectations for AI tools to assist them by, for instance, alleviating some of the physical labor involved (e.g., helping them read stories), generating valuable and high-quality interactive content or questions for them when they lack expertise, providing inspiration for interactions with their children, and preparing appropriate resources for supporting their storytelling with children. These supporting functions could be considered by future AI-based interactive storytelling tools.

\textbf{\textit{Taking Parents as Beneficiaries: Beyond a Storytelling Toolman.}} With taking parents as beneficiaries, we also suggest shifting parents’ role in AI-based storytelling from merely tool man or technology supporters, as most existing tools do, to active participants. 
Activities such as role-playing and interactive play can be incorporated. While role-playing is already utilized in some tools \cite{luka, storycoder}, many of them still take parents as supporters in terms of helping with media content selection and text-to-speech usage, rather than fully participating as players. We recommend treating parents as true players and enhancing their involvement. Ongoing research on interactive technologies (e.g., \cite{chen2022designing}) may offer potential support for better parent-child interactive storytelling experiences.

\textbf{\textit{Enabling Parents' Personal Experience Through Co-Creating.}} 
According to most of our participants, during their actual storytelling with preschoolers, their conversations often go beyond the storybooks, and integrate a lot of personal viewpoints and experiences, which has also been recognized in the literature as an essential and valuable way of storytelling \cite{parentrole2, Storybuddy}. We therefore 
propose AI-based storytelling tools could empower parents by involving them as co-storytellers (i.e., human-AI collaboration~\cite{wang2020human}), and providing them interfaces to seamlessly incorporate their personal experiences, childhood memories, or whatever they want into the narrative. Such integration could also effectively contribute to the realization of personalized story themes that align with parental objectives.

\subsection{Design Considerations for Child-Appropriated Interaction: Enabling Preschoolers' Psychology and Cognitive Features}
Our study has demonstrated that preschoolers often face certain cognitive difficulties in accepting the virtual interactive modality; Meanwhile, current AI tools often fail to effectively support preschoolers' proactive and consistent participation. The main reason for these issues is that the current designs do not take into account the preschoolers' cognitive and psychological features, resulting in a misalignment between child cognitive development and the virtual interactive way~\cite{xu2020you}.

\CSCWreview{According to these findings, we propose that future AI storytelling tools need to fit with preschoolers' psychology and cognitive characteristics to make children be a more active role in storytelling processes.}
Some classical theories of child cognitive development, such as Piaget's theory of cognitive development \cite{piaget1976piaget}, could be used to guide the interactive design tailored for preschoolers. 
Within the framework of Piaget's theory, each stage of development has distinct cognitive characteristics. Preschoolers are in the preoperational stage, where their cognitive processes remain anchored in concrete experiences, starting to be symbolic thinking. They can use imagination and symbols to express and solve problems, but their grasp of logical reasoning and abstract concepts is still limited\cite{piaget1976piaget}.


Given the cognitive and psychological feathers of preschoolers presented by \citet{piaget1976piaget}, as well as our findings, we propose concrete, multimodal, and tangible interaction modalities for future AI-based storytelling tools for preschoolers, such as providing tangible user interfaces (e.g., physical objects) for virtual interactions, incorporating motion perception, to enhance user engagement; Supporting storytelling processes with concrete experiences (e.g. AI building blocks \cite{cai2023emotionblock}) to capture children's attention; Enabling preschoolers to engage in storytelling through a combination of real and virtual interactions (e.g. creating 3D story environments where children can manipulate virtual characters physically to accomplish story narrations \cite{wang2014storycube}).
These approaches provide a foundation for concrete thinking in children at the preoperational stage and, with the aid of virtual interaction modalities, facilitate the development of symbolic thinking.

It is noteworthy that we are not denying the value of virtual interaction modalities. Instead, similar to existing studies \cite{xu2021same, pantoja2019voice}, our participants also reported benefits of virtual (e.g. voice-based) interaction in enhancing children's language abilities. 
What we want to emphasize here is the importance of combining multiple interaction modalities. That is, integrating virtual interaction modality into tangible interfaces, serving as a medium for children to transition between concrete thinking and symbolic thinking.

\subsection{Design Considerations for Child-Appropriated AI Algorithms: Carefully Considering "Appropriated"}
In terms of core technologies and algorithms of existing AI-based storytelling tools, our participants reported significant challenges, primarily regarding the silly, unintelligent algorithms, and their failure to output adaptive and understandable content to children and answer their "a hundred thousand whys". From the technical perspective, a natural improving suggestion is to construct child-appropriated datasets and corpora that cater to preschoolers' language and cognitive abilities, and then use them to enhance algorithm performance, making the output better suited to preschoolers' needs in terms of language style, content, complexity, etc. Ongoing efforts have started to focus on these aspects, such as constructing question-answer storytelling datasets tailored to children's needs \cite{FairytaleQA, chen2023fairytalecqa}, utilizing Supervised Fine-Tuning (SFT~\cite{xu2024mental}) or Knowledge Graph-Augmented Generation approaches (KG-RAG) for knowledge expansion \cite{chen2023fairytalecqa}, and improving AI to generate story book related, adaptive conversation \cite{yao2022ai, kory2014storytelling}. 
We believe that these research directions are crucial for the future AI-based storytelling technologies.

Moving forward, we further suggest carefully considering what is "appropriate" for preschoolers and, based on that, exploring directions for algorithmic improvement. 
When we refer to ``children-appropriated'', we are considering whether the interaction and generated content are suitable for children. This entails aligning with their cognitive level, personal abilities, interests, preferences, and other relevant factors \cite{moyer1987child}. 
However, from the algorithmic perspective, the ``appropriate'' is often reflected in the technical performance, e.g. the accuracy of answers, which the majority of our participants believe may not be applicable for preschoolers, because there is no ``standard or correct answer'' for each child. 
For instance, algorithms may focus on whether it can generate the correct answer ``\textit{poisoned apple}'' for the question ``\textit{what did Snow White eat and die from?}'', while children may be more interested in questions like ``\textit{What dress was the witch wearing?}''
We therefore suggest future algorithms to support more open and diverse interactions that encourage children's creative and critical thinking, rather than solely focusing on providing the "accurate" answer. The advancing large language models have the potential to support such conversational needs, and researchers have also started to examine its potential in supporting children’s creative expression and storytelling \cite{han2023design, AIGC2023design}. We believe it is a valuable direction, but need to carefully consider content safety and ethical issues which are significant concerns raised by our participants.

To conclude our discussion, we emphasize that not all challenges raised by parents can or should be resolved through AI algorithms, because, given AI's understanding of ``appropriate'', it might be insufficient to answer the countless ``why'' questions from preschoolers that have no correct answer, and support parents expected storytelling and personalized interactions. Further, 
AI solutions will inevitably involve the use of more perceptual technologies to gather additional data, posing challenges to the protection of children's privacy. Our participation also expresses similar concerns in this regard.
 Echoing the discussion in section 5.1, we further emphasize it is crucial to carefully consider the role and functionality of AI tools in preschoolers' storytelling - these tools should serve as aids for parents, enabling them to better support the storytelling process in a manner that aligns with their children's unique needs.

\subsection{Limitation and Future Work}
The major limitation of our study is the small sample size, which may result in possible sample selection bias \cite{heckman1979sample} and sample diversity~\cite{wolda1981similarity} issues. In particular, during the recruitment process, we did not consider the diverse characteristics of parents. However, during the research, we discovered that participants' individual characteristics, such as their economic level, educational background, and professional expertise, significantly influenced their attitudes and perceptions toward using AI for storytelling. Yet, given the sample size of this study, we did not analyze these perspectives very well.
\CSCWreview{Meanwhile, we recruited parents who had already inclined to use AI-based products. While such kind of participants could provide us with richer usage experiences with AI-based tools, the attitudes and perceptions of parents who have not used AI-based products might be ignored. In addition, all our participants were recruited from China, which might limit the generalizability of our findings.}


In future research, we aim to expand our focus to a more diverse range of parental perspectives, painting a more complete picture of our research questions.
\CSCWreview{Meanwhile, we will involve more stakeholders, including children and kindergarteners, to the study, and identify their practical needs and experiences, through which to shed light on future human-centered design in AI-based storytelling scenarios for preschoolers. 
}

\section{Conclusion}
Using AI-based tools to support parent-child interactive storytelling has been a contemporary trend. In this work, we conducted an interview study to deeply examine how parents use, experience, and perceive AI-based storytelling technologies in practical storytelling scenarios. 
Our findings suggest current AI-based storytelling technologies face numerous challenges in supporting storytelling for preschoolers as parents expect.
We categorize them into interactive and algorithmic challenges, and elaborate on them.
Drawing on these findings, we identified vital design implications for AI technologies to better facilitate children-centered interactive storytelling.

With the widespread applications of AI in contemporary society, human-centered AI has become a focal point in CSCW, HCI and AI fields. However, how to effectively implement the human-centered vision and create practical technologies that truly put people first has consistently posed challenges.
Our research focuses on the specific AI application scenario – AI-based interactive storytelling for preschoolers, and delves into how the concept of human-centered AI can be realized in this scenario. While our research focuses on the specific scenario of storytelling, we believe that the design principles we discuss can be generalized to a broader range of technology designs aiming at facilitating children-AI interaction.



\bibliographystyle{ACM-Reference-Format}
\bibliography{sample-base}









\end{document}